\documentclass[12pt,5p]{elsarticle}

\usepackage{graphicx}
\graphicspath{{Plots/}}

\usepackage{amsmath}
\usepackage{amssymb}
\usepackage{url}

\def\units#1{\hbox{$\,{\rm #1}$}}

\journal{Astroparticle Physics}

\begin{document}

\begin{frontmatter}

\title{A simple Local Interstellar Spectrum model to fit the proton fluxes measured by the AMS and PAMELA detectors}

\author[a,b]{F.~Loparco}
\author[a]{M.~N.~Mazziotta\corref{cor1}}
\cortext[cor1]{Corresponding authors: Tel.: +390805443163 (M.N.~Mazziotta), fax +390805442470, Istituto Nazionale di Fisica Nucleare, Sezione di Bari, Via Orabona, 4 - 70125 Bari, Italy}
\ead{mazziotta@ba.infn.it}

\address[a]{Istituto Nazionale di Fisica Nucleare, Sezione di Bari, 70125 Bari, Italy}
\address[b]{Dipartimento di Fisica ``M. Merlin" dell'Universit\`a e del Politecnico di Bari, I-70126 Bari, Italy}

\begin{abstract}
In this paper we discuss some simple analytical models to fit the cosmic-ray (CR) proton data collected by 
the AMS detector in June 1998 and by the PAMELA detector in several campaigns covering the period 2006-2009. 
The CR proton spectrum at Earth is derived starting from the model of the local interstellar spectrum (LIS) 
and folding it with the solar modulation potential in the force field approximation.
The data are well described by a LIS modeled with a simple power law particle momentum density.   
\end{abstract}


\end{frontmatter}

\section{Introduction}

Cosmic rays (CRs) interact with gas atoms during their propagation in the interstellar medium and can suffer
significant energy losses, thus modifying their injection spectra and composition. In addition, the spectra 
of CRs reaching the Earth are affected by the solar wind and the by the solar magnetic field (solar modulation effect). 
The solar modulation plays a relevant role on CR spectra in the low energy region, and its effect needs to 
be disentangled to allow a comprehensive picture to emerge. In fact, to understand the origin and propagation 
of CRs, a knowledge of their energy spectra in the interstellar medium is required. 

Precise measurements of CR spectra over a wide rigidity range, from a few hundred $\units{MV}$ to tens 
of $\units{GV}$ can be used to study the effect of solar modulation, including 
the convective and adiabatic cooling effect of the expanding solar wind and the diffusive and particle 
drift effects of the turbulent heliospheric magnetic field (HMF).

A full three-dimensional (3D) model was developed to compute the differential intensity of CR protons from 
$10~\units{MeV}$ to $30~\units{GeV}$ at Earth~\cite{potgieter2013}, and was applied to give an interpretation 
of the PAMELA proton data sets collected from 2006 to 2009~\cite{pamelatime}. The model also includes a 
detailed treatment of the CR propagation in the solar magnetic field, that allows a precise description of 
the solar modulation effect.  
The implementation of this approach requires to provide the local interstellar proton spectrum (LIS) 
as initial condition. The LIS input spectrum is then  ``modulated'' by the solar magnetic field,
that affects the shape of the CR spectrum at Earth.

The choice of the LIS has always been rather contentious (see for instance \cite{igor2002}), and its 
parametrization in terms of proton kinetic energy could be complex (see for instance ~\cite{potgieter2013}, 
\cite{webber2010}, \cite{igortroy}). In this paper we assume some simple analytical LIS models to fit the 
proton fluxes measured by the AMS detector in June 1998~\cite{ams98} and by the PAMELA detector in different periods 
from 2006 to 2009. The solar modulation effect is also described in a simple form, using the force-field 
approximation~\cite{gleeson}.

\section{Local proton spectrum models}

The simplest model describing CR acceleration is the first-order Fermi mechanism, where particles gain
energy by diffusing back and forth across a shock front while convecting downstream.
In this framework, the particle differential density per unit momentum $n(p)$ 
is $\propto p^{-a}$. After injection 
into the interstellar medium with spectral index $a \simeq 2.1 \div 2.2$, characteristic 
of supernova remnant (SNR) shocks, CRs are transported in the astronomical environments with 
rigidity-dependent escape lengths, that soften their spectra by $\delta \simeq 0.5 \div 0.6$, 
leaving a steady-state CR particle momentum density $n(p) \propto p^{-\alpha}$, 
where $\alpha = a + \delta \simeq 2.7 \div 2.8$~\cite{dermer2012}.

\begin{figure}[!ht]
\includegraphics[width=0.9\columnwidth,keepaspectratio,clip]{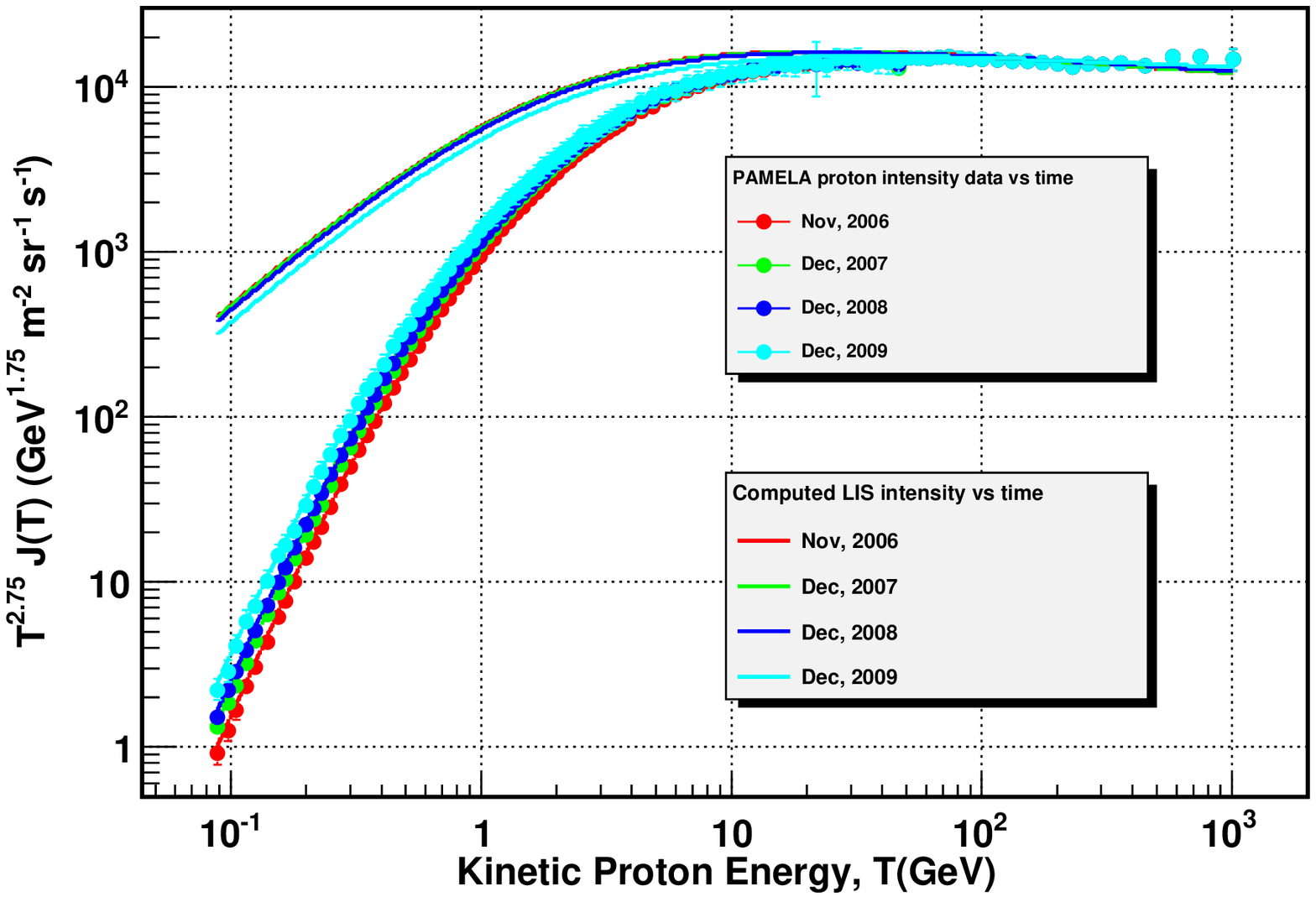}
\includegraphics[width=0.9\columnwidth,keepaspectratio,clip]{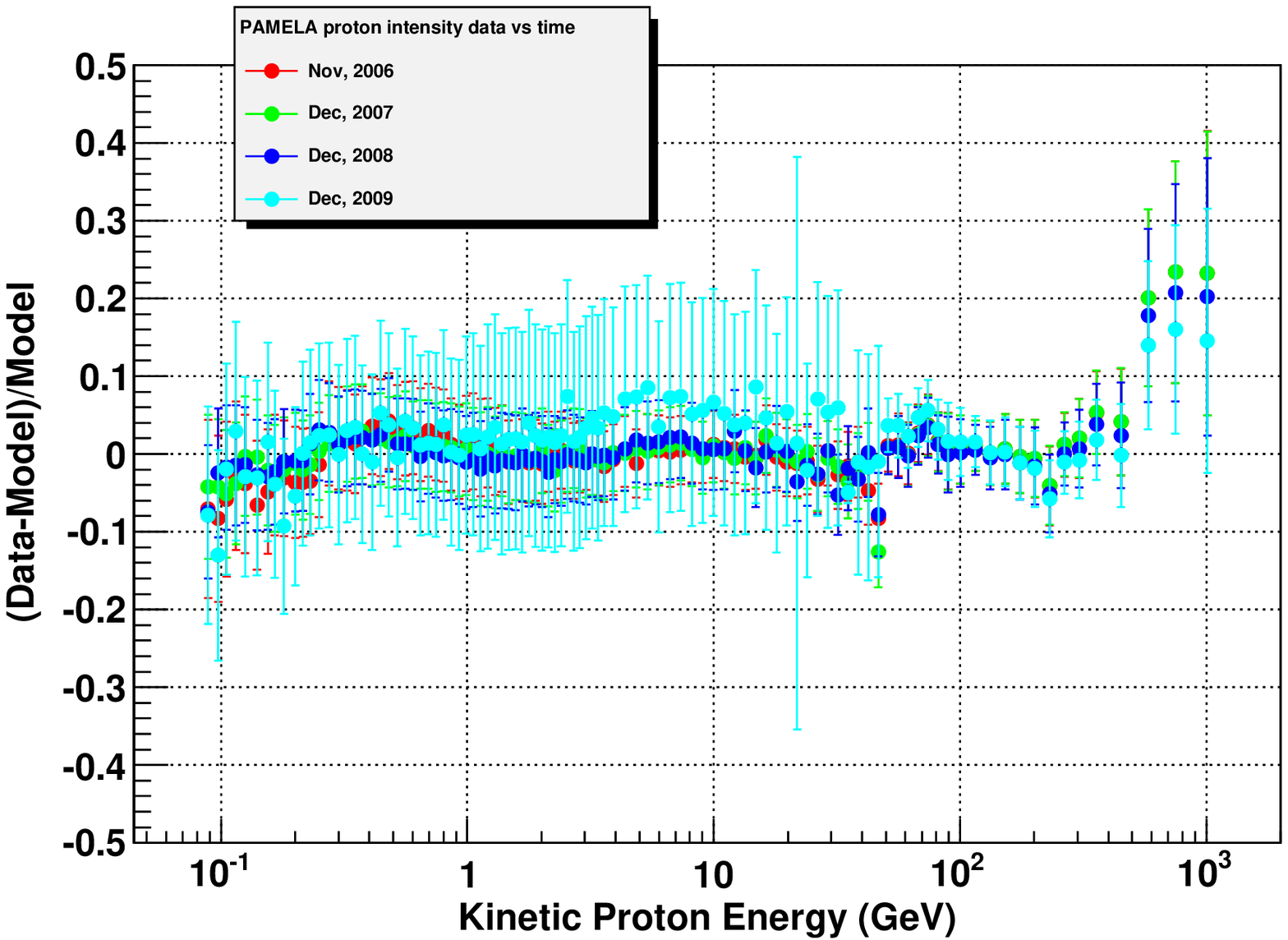}
\caption{Fit of the PAMELA proton data collected in November 2006 (red), December 2007 (green), December 2008 (blue)
and December 2009 (cyan) with a simple momentum power law LIS folded with solar modulation (the fit was performed up to $175~\units{GeV}$). The top panel shows the fit results superimposed to the the data points. The dashed lines show 
the fitted spectra (Eq.~\ref{eqearth}) and the continuous lines show the corresponding LIS (Eq.~\ref{eqlis}). 
The bottom panel shows the fit residuals.}
\label{Fig1}
\end{figure}

\begin{figure}[!ht]
\includegraphics[width=0.9\columnwidth,keepaspectratio,clip]{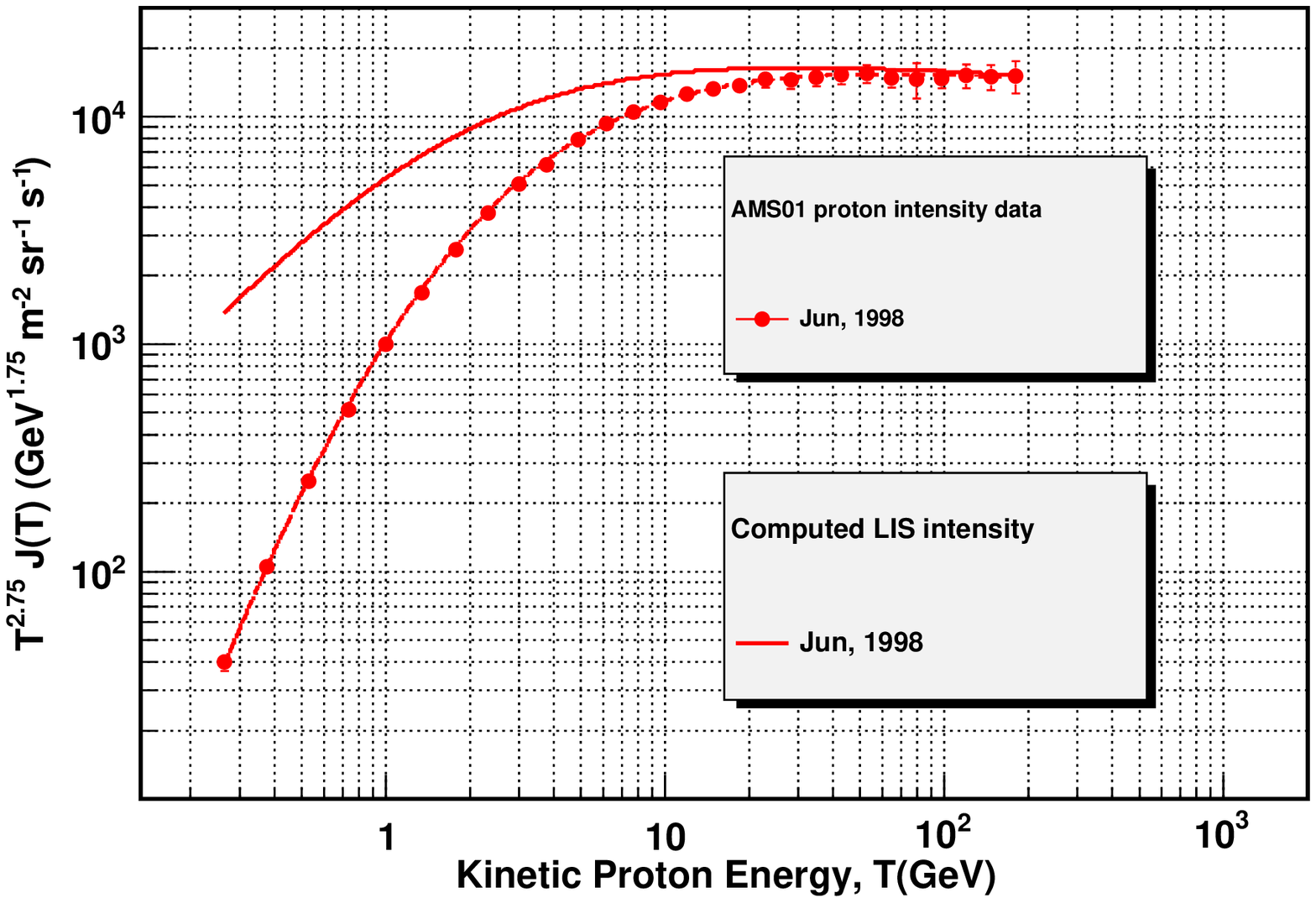}
\includegraphics[width=0.9\columnwidth,keepaspectratio,clip]{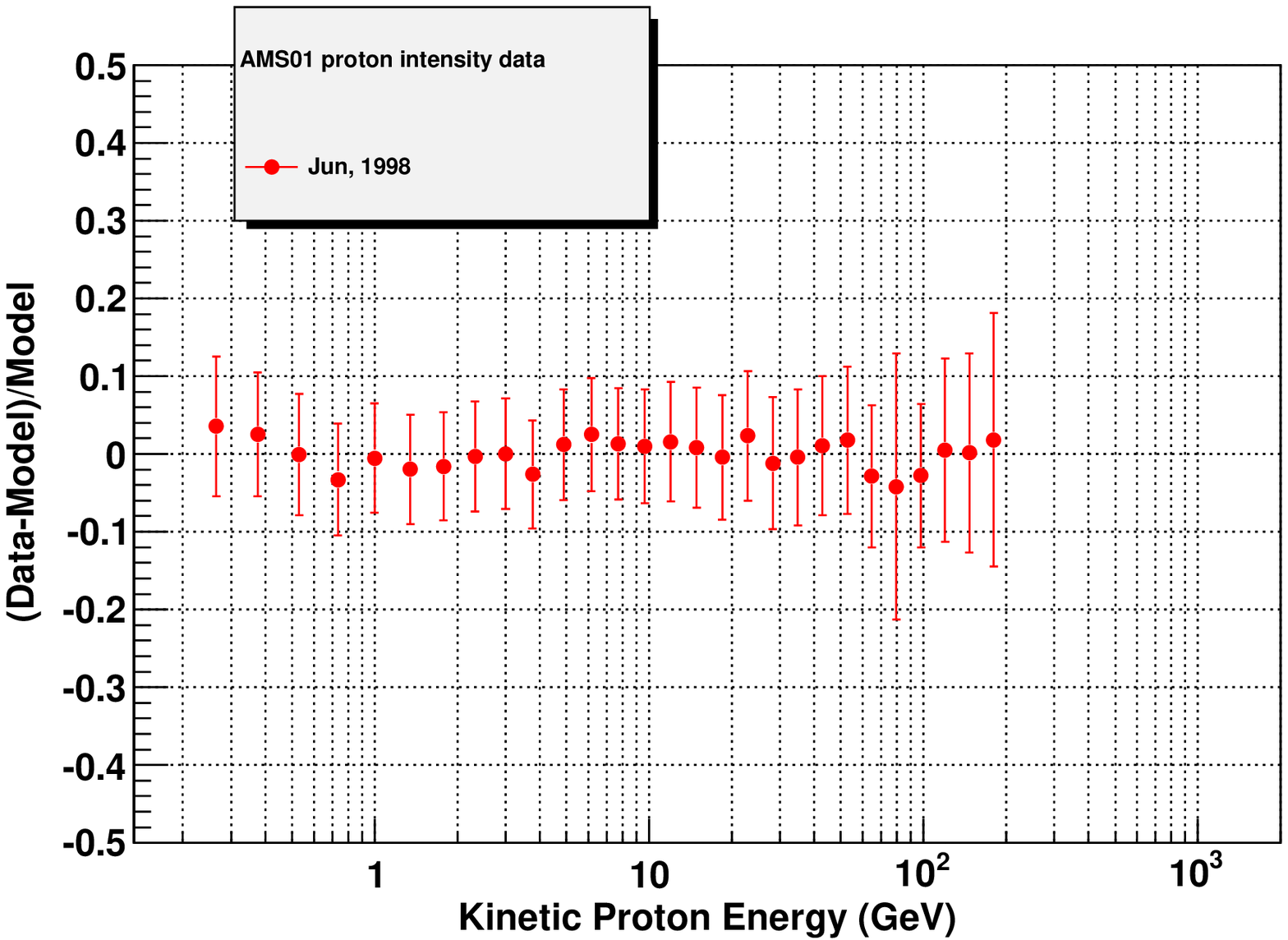}
\caption{Fit of the AMS proton data collected in June 1998 (red) with a simple momentum power law LIS folded 
with solar modulation. The top panel shows the fit results superimposed to the the data points. The dashed lines show 
the fitted spectra (Eq.~\ref{eqearth}) and the continuous lines show the corresponding LIS (Eq.~\ref{eqlis}). 
The bottom panel shows the fit residuals vs energy. }
\label{Fig2}
\end{figure}

Therefore, in the present work we will consider a simple LIS model with the 
differential particle momentum density in the form:

\begin{equation}
 n(p) = k_{0} \left( \frac{p}{p_{0}} \right)^{-\alpha}.
\end{equation}
The spectral differential intensity in momentum is obtained 
by multiplying $n(p)$ for the factor $\beta c / 4\pi$, where $\beta c$ 
is the particle velocity: 

\begin{equation}
J(p) = \frac{\beta c}{4\pi} n(p) = k \beta \left( \frac{p}{p_{0}} \right)^{-\alpha}.
\label{eqlispowerlaw}
\end{equation}
where $k=k_{0}c/4\pi$. Hereafter we will assume $c=1$ and we will 
express both energies and momenta in units of $\units{GeV}$. In writing 
the previous equations we introduced a momentum scale $p_{0}=1~\units{GeV}$.
Therefore $k_{0}$ will be expressed in the same units as $n(p)$, i.e. in
$\units{GeV^{-1} m^{-3}}$ and $k$ will be expressed in the
same units as $J(p)$, i.e. in $\units{GeV^{-1} m^{-2} s^{-1} sr^{-1}}$. 

The differential intensity in momentum can be converted into a differential intensity 
in kinetic energy taking into account that:

\begin{equation}
J(T) = \frac{dp}{dT}~J(p)
\end{equation}
\begin{equation}
\frac{dp}{dT} = \frac{T+m}{\sqrt{T(T+2m)}} = \frac{1}{\beta}
\end{equation}
where $m$ is the particle rest mass. 
The CR spectrum in the interstellar space as a function of the kinetic energy is therefore given by:

\begin{equation}
J_{LIS}(T) = k \left[ \frac{T(T+2 m)}{p_{0}^{2}} \right]^{-\frac{\alpha}{2}}
\label{eqlis}
\end{equation}
 
The spectral index in kinetic energy, $\alpha_T$, is defined as:
\begin{equation}
\alpha_T = -\cfrac{d \log J_{LIS}(T) }{d \log T} = \alpha \cfrac{T+m}{T+2m}. 
\end{equation}
The previous result shows that the CR spectrum exhibits a change of curvature with increasing 
kinetic energy ($\alpha_{T} \rightarrow \alpha/2$ for $T \rightarrow 0$ while 
$\alpha_{T} \rightarrow \alpha$ for $T \rightarrow \infty$).

\begin{figure}[!ht]
\includegraphics[width=0.9\columnwidth,keepaspectratio,clip]{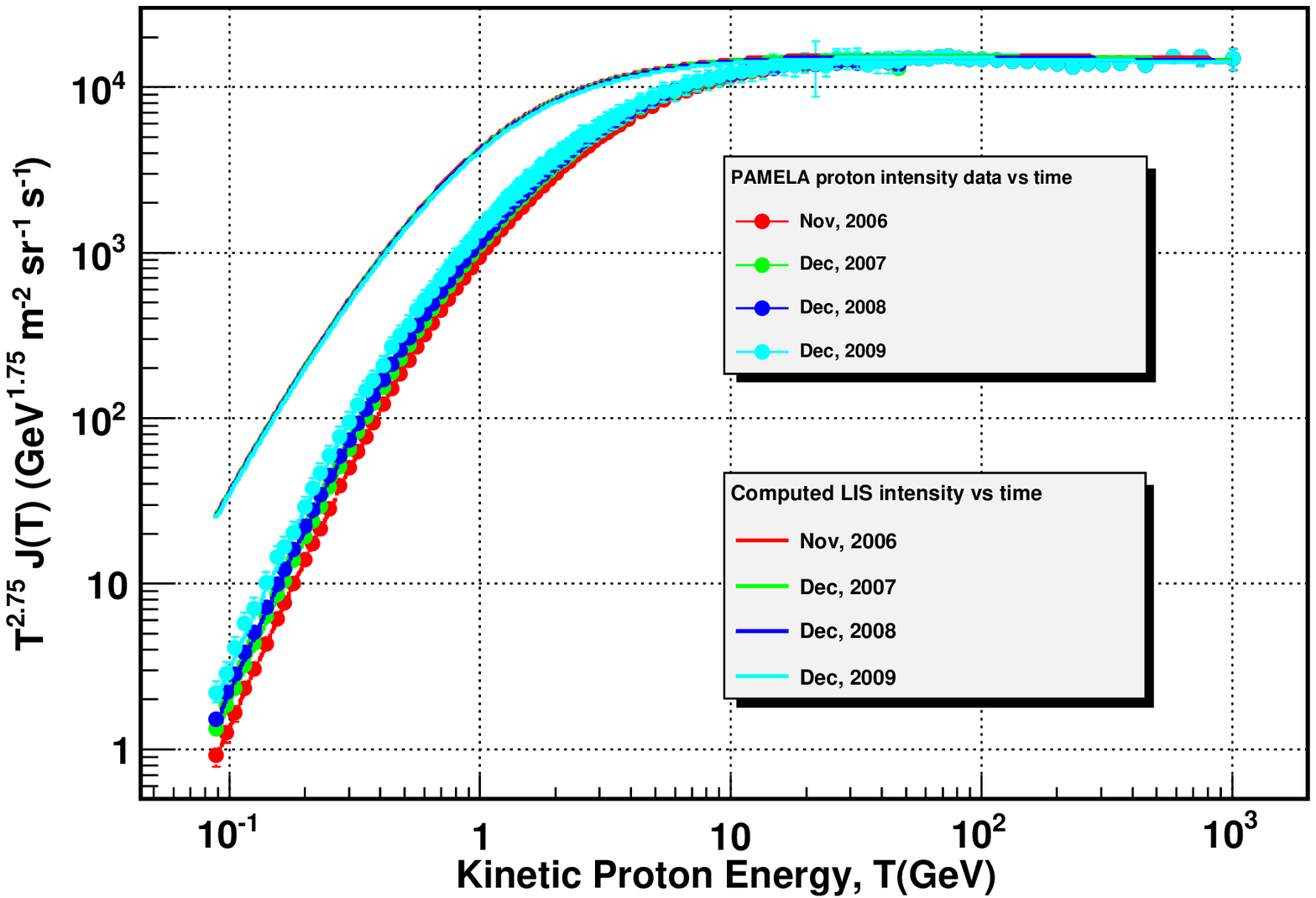}
\includegraphics[width=0.9\columnwidth,keepaspectratio,clip]{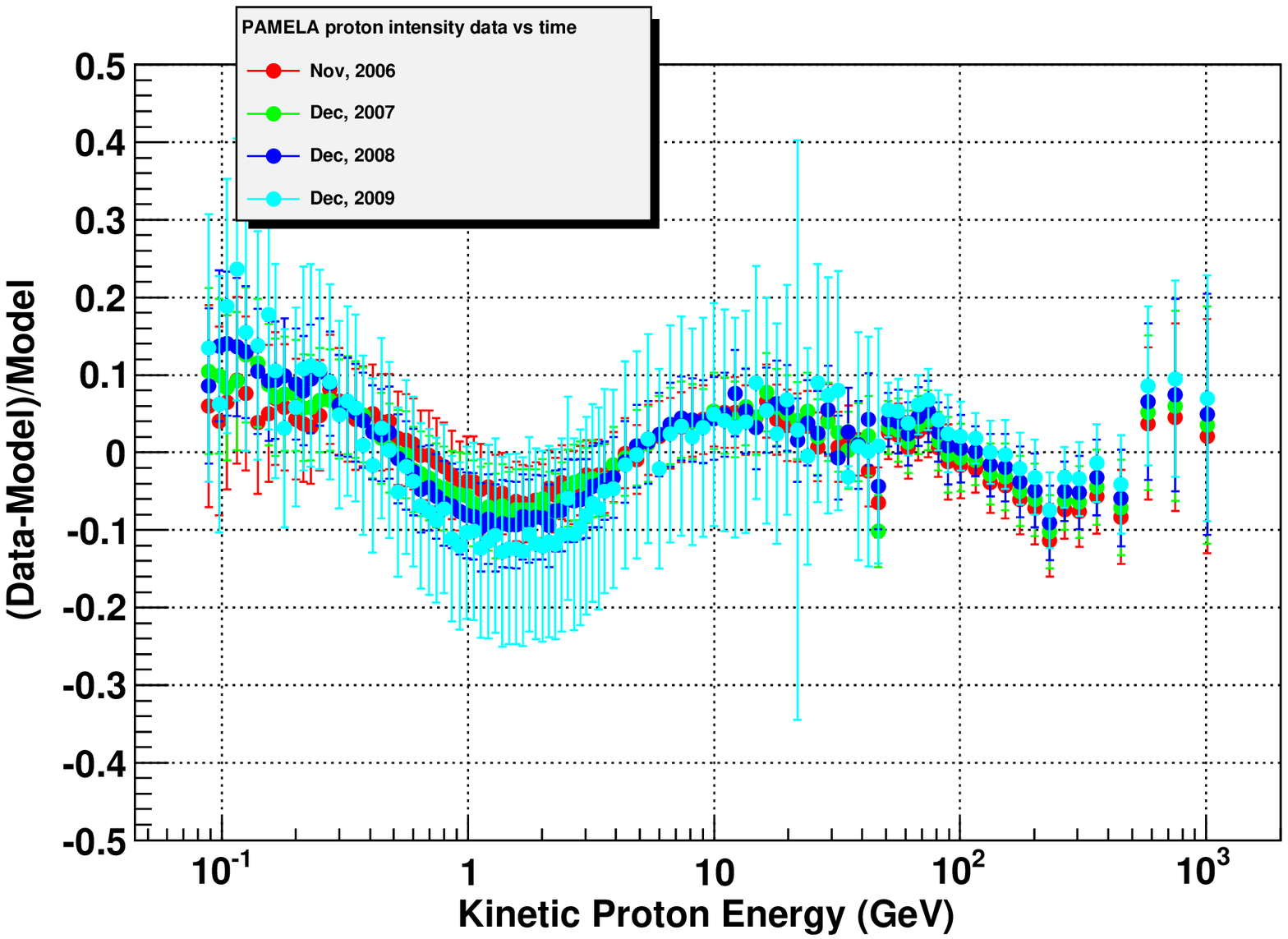}
\caption{Fit of the PAMELA proton data collected in November 2006 (red), December 2007 (green), December 2008 (blue)
and December 2009 (cyan) with the LIS of Eq.~\ref{eqlispam} folded with solar modulation (the fit was 
performed up to $175~\units{GeV}$). The top panel shows the fit results superimposed to the the data points. 
The dashed lines show the fitted spectra and the continuous lines show the corresponding LIS (Eq.~\ref{eqlispam}). 
The bottom panel shows the fit residuals.}
\label{Fig3}
\end{figure}

The LIS model of Eq.~\ref{eqlis} can be folded with the solar modulation effect in the force field approximation
introducing the solar modulation potential field $\Phi$, and results into a CR spectrum at Earth given by:
\begin{equation}
J_{Earth}(T) = J_{LIS}(T+\Phi Z/A) \frac{T(T+2m)}{\left(T+m+\Phi Z/A\right)^2-m^2}
\label{eqearth}
\end{equation}
where $Z$ and $A$ are the atomic and mass number respectively of the given CR species.

Figure~\ref{Fig1} shows the results obtained fitting the PAMELA proton data with Eq.~\ref{eqearth}. 
The data up to $48~\units{GeV}$ have been taken from Table 1 of  Ref.~\cite{pamelatime}. 
The data at higher energies, from $48~\units{GeV}$ to $1~\units{TeV}$ have been taken from Ref.~\cite{pamelascience}. 
The fit was performed up to $175~\units{GeV}$ using the MINUIT package implemented in the ROOT framework~\cite{ROOT}. 
The residuals exhibit very small fluctuations, within a few $\%$. It is worth to point out that the 2009 data show 
some large point-to-point fluctations, probably due to the fact that these data were collected at the 
end of the $11^{th}$ Solar cycle. 
Figure~\ref{Fig2} shows the results obtained fitting the AMS proton data with Eq.~\ref{eqearth}. 
The data have been taken from Table 3 of Ref.~\cite{ams98}. The error bars shown in the plots are evaluated 
by adding in quadrature statistical and systematic uncertainties. Also in this case the residuals 
exhibit small fluctuations.

The results of all the fits are summarized in Table~\ref{tabfit}. It is interesting to point out that
both the AMS and PAMELA data sets are well fitted by the momentum power law LIS. Moreover, the values of 
the LIS parameters (prefactor and spectral index) obtained from the fit of the AMS 
data are consistent with those obtained from the fits of the PAMELA data sets taken in 2006, 2007 and
2008, as one would expect since the LIS is time independent. On the other hand, the fit of the data
collected by PAMELA in 2009 yields values that differ significantly from those obtained 
from the other fits.  
Because of the consistency between the AMS and the first three PAMELA fits, we decided to combine 
these results into a unique LIS, with a spectral index $\alpha=2.853$ and a pre-factor 
of $2.4863 \times 10^4~\units{GeV^{-1} m^{-2} s^{-1} sr^{-1}}$, evaluated from the weighted average of
the individual fit results. 
Using these parameters for the LIS (Eq.~\ref{eqlis}) and leaving only the solar modulation potential free, 
the fits of the AMS and of the PAMELA data do not change significantly with respect to
the ones shown in Figures~\ref{Fig1} and ~\ref{Fig2}. In particular, the fitted values of the
solar modulation potential do not exhibit significant variations (see Table~\ref{tabfit}).

We have also fitted the data samples using the LIS model given in Ref.~\cite{potgieter2013}, i.e.:
\begin{equation}
J_{LIS}(T) =
\left\{ 
\begin{array}{ll}
707~k~ \textrm{e}^{ 4.64 -0.08 (\log T)^{2} -2.91 \sqrt{T}  }& T<T_{0} \\
& \\
685~k~ \textrm{e}^{ 3.22 -2.78 \log T - 1.5/T } & T>T_{0} \\
\end{array}
\right.
 \label{eqlispam}
\end{equation}
where $T_{0}=1.4~\units{GeV}$. The numerical coefficients in the previous equation include
units of measurement: $J_{LIS}(T)$ is given in units of $\units{GeV^{-1} m^{-2} s^{-1} sr^{-1}}$
if $T$ is expressed in $\units{GeV}$. The coefficient $k$ is a scale factor 
($k=1$ will reproduce exactly the formula in Eq.~13 of Ref.~\cite{potgieter2013}).

Figure~\ref{Fig3} shows the results of the fits performed assuming the LIS model in Eq.~\ref{eqlispam} folded 
with the solar modulation in the force field approximation. In this case the residuals show larger fluctuations
with respect to the simple power law fits and the $\chi^{2}$ values are higher. The fit results are 
summarized in Table~\ref{tabfit}. It is also worth to point out that, when this LIS model is assumed, 
the solar modulation potential values are smaller with respect to those obtained assuming the simple 
momentum power law model. This feature is due to the shape of the LIS given in Eq.~\ref{eqlispam}, 
that predicts a curvature at low energies. Another interesting feature of these fits is the value of 
the prefactor, that in all cases is consistent with $1$, as expected.
We have also performed the fit of the AMS and PAMELA data with the LIS model of Eq.~\ref{eqlispam} 
with a fixed prefactor $k=1$. This constraint does not worsen the fit, and the  
the solar modulation potential values do not change significantly. 

To reproduce the B/C ratio the injection spectrum of CR protons is often approximated with a power law in the 
rigidity space with a break at a few $\units{GV}$(see for instance ~\cite{igor2002}). A possible description of 
this feature can be given by choosing for the LIS momentum density a broken power law function 
(with a discontinuity in the first derivative at the break):

\begin{equation}
n_{LIS}(p) =
\left\{ 
\begin{array}{ll}
k_{0}~ \left(\frac{p}{p_b}\right)^{-\alpha_1}  &  p<p_{b} \\
& \\
k_{0}~ \left(\frac{p}{p_b}\right)^{-\alpha_2}  &  p\geq p_{b} \\
\end{array}
\right.
\end{equation}
that result into a differential intensity given by:

\begin{equation}
J_{LIS}(T) =
\left\{ 
\begin{array}{ll}
k~ \left(\frac{p}{p_b}\right)^{-\alpha_1}  &  p<p_{b} \\
& \\
k~ \left(\frac{p}{p_b}\right)^{-\alpha_2}  &  p\geq p_{b} \\
\end{array}
\right.
 \label{eqlisbb}
\end{equation}
with $p=\sqrt{T(T+2 m)}$ and $k=k_{0}c/4\pi$. The break momentum $p_{b}$ corresponds to
a break kinetic energy $T_{b} = \sqrt{p_{b}^{2}+m^{2}}-m$. 

The results of the fits with the LIS of Eq.~\ref{eqlisbb} are also shown in Table~\ref{tabfit}.
We note that the fits yield no evidence of a break, since $\alpha_{1} \approx \alpha_{2}$ 
within the errors for all the data sets.

\section{Conclusions}
The simple power law model of the proton LIS folded with the solar modulation in the force field 
approximation provides a good fit of the both AMS and PAMELA proton data. 
This result needs to be investigated with further analyses, for instance by using data 
from helium and heavy nuclei on short time periods, that are not publicly available.

In general, it is worth to point out that a CR measurement at Earth does not allow to easily 
reconstruct the LIS, since the Solar modulation effect cannot be easily disentangled.
However, a constraint to the LIS spectrum could be provided by a fit of the gamma-ray 
emissivity of the local neutral gas measured by the Fermi LAT~\cite{dermer2012,latemiss,jeanmarc}. 

\section*{Acknowledgements}

We are grateful to Charles D. Dermer and Andrew W. Strong for the fruitful discussion during the preparation of 
the manuscript and for their valuable contribution. 

\begin{center}

\begin{table*}[h!]

\tiny

\begin{tabular}{|l|c|c|c|c|c|}
\hline
\hline
\multicolumn{6}{|c|}{Simple power law LIS (Eq.~\ref{eqlis})} \\
\hline
\hline
 & AMS Jun 1998 & PAMELA Nov 2006 & PAMELA Dec 2007 & PAMELA Dec 2008 & PAMELA Dec 2009 \\
 \hline
 $k (10^{4}~\units{GeV^{-1}m^{-2}s^{-1}sr^{-1}})$ & $2.342 \pm 0.193$ & $2.548 \pm 0.067$ 
 & $2.520 \pm 0.065$ & $2.414 \pm 0.064$ & $2.072 \pm 0.097 $\\ 
 \hline
 $\alpha$ & $2.832 \pm 0.024$ & $2.858 \pm 0.007$ &$2.857 \pm 0.007$ &$2.847 \pm 0.007$ &$2.818 \pm 0.011$ \\
  \hline
 $\Phi (\units{GeV})$ & $0.629 \pm 0.030$ & $0.685 \pm 0.009$ &$0.602 \pm 0.008$ &$0.557 \pm 0.008$ &$0.454 \pm 0.012$ \\
 \hline
 $\chi^{2}/d.o.f.$ & $1.4/25$ & $16.2/87$ &$15.1/87$ &$11.1/87$ &$28.2/87$ \\
 \hline
 \hline
\end{tabular}

\vspace{0.5cm}

\begin{tabular}{|l|c|c|c|c|c|}
\hline
\hline
\multicolumn{6}{|c|}{Simple power law LIS (Eq.~\ref{eqlis}) with $k=2.4863\times 10^{4}~\units{GeV^{-1}m^{-2}s^{-1}sr^{-1}}$ and $\alpha=2.853$} \\
\hline
\hline
 & AMS Jun 1998 & PAMELA Nov 2006 & PAMELA Dec 2007 & PAMELA Dec 2008 & PAMELA Dec 2009 \\
 \hline
 $\Phi (\units{GeV})$ & $0.646 \pm 0.010$ & $0.676 \pm 0.003$ &$0.598 \pm 0.003$ &$0.566 \pm 0.003$ &$0.501 \pm 0.005$ \\
 \hline
 $\chi^{2}/d.o.f.$ & $2.4/27$ & $17.8/89$ &$15.5/89$ &$12.5/89$ &$59.3/89$ \\
 \hline
 \hline
\end{tabular}

\vspace{0.5cm}

\begin{tabular}{|l|c|c|c|c|c|}
\hline
\hline
\multicolumn{6}{|c|}{LIS from ref.~\cite{potgieter2013} (Eq.~\ref{eqlispam})} \\
\hline
\hline
 & AMS Jun 1998 & PAMELA Nov 2006 & PAMELA Dec 2007 & PAMELA Dec 2008 & PAMELA Dec 2009 \\
 \hline
 $k$ & $1.062 \pm 0.025$ & $1.044 \pm 0.078$ & $1.029 \pm 0.080$ & $1.015 \pm 0.080$ & $0.995 \pm 0.089$ \\
 \hline
 $\Phi (\units{GeV})$ & $0.583 \pm 0.018$ & $0.594 \pm 0.005$ & $0.500 \pm 0.005$ & $0.459 \pm 0.005$ & $0.377 \pm 0.007$ \\
 \hline
 $\chi^{2}/d.o.f.$ & $11.6/26$ & $56.9/88$ & $82.3/88$ & $104.4/88$ & $54.1/88$ \\
 \hline
 \hline
\end{tabular}

\vspace{0.5cm}

\begin{tabular}{|l|c|c|c|c|c|}
\hline
\hline
\multicolumn{6}{|c|}{LIS from ref.~\cite{potgieter2013} (Eq.~\ref{eqlispam}) with $k=1$} \\
\hline
\hline
 & AMS Jun 1998 & PAMELA Nov 2006 & PAMELA Dec 2007 & PAMELA Dec 2008 & PAMELA Dec 2009 \\
 \hline
 $\Phi (\units{GeV})$ & $0.550 \pm 0.011$ & $0.572 \pm 0.004$ & $0.488 \pm 0.003$ & $0.453 \pm 0.003$ & $0.379 \pm 0.007$ \\
 \hline
 $\chi^{2}/d.o.f.$ & $17.9/27$ & $88.3/89$ & $95.3/89$ & $107.6/89$ & $54.4/89$ \\
 \hline
 \hline
\end{tabular}

\vspace{0.5cm}

\begin{tabular}{|l|c|c|c|c|c|}
\hline
\hline
\multicolumn{6}{|c|}{Broken power law LIS (Eq.~\ref{eqlisbb})} \\
\hline
\hline
 & AMS Jun 1998 & PAMELA Nov 2006 & PAMELA Dec 2007 & PAMELA Dec 2008 & PAMELA Dec 2009 \\  \hline
$k (\units{GeV^{-1}m^{-2}s^{-1}sr^{-1}})$ & $134.38 \pm 25.18$& $157.24 \pm 16.30$ &$155.69 \pm 10.14$ &$153.36 \pm 10.42$ &$132.67 \pm 29.38$ \\  \hline
$\alpha_1$ & $2.705 \pm 0.155$& $2.942 \pm 0.089$ &$2.888 \pm 0.053$ &$2.844 \pm 0.051$ &$2.908 \pm 0.112$ \\  \hline
$\alpha_2$ &$2.824 \pm 0.024$ &$2.863 \pm 0.012$ &$2.858 \pm 0.007$ &$2.847 \pm 0.007$ &$2.817 \pm 0.012$ \\  \hline
$\Phi (\units{GeV})$ & $0.566 \pm 0.080$ & $0.722 \pm 0.039$ &$0.615 \pm 0.023$ &$0.556 \pm 0.021$ &$0.484 \pm 0.043$ \\  \hline
$p_{b} (\units{GeV})$ &$6.13 \pm 0.41$ & $5.96 \pm 2.11$ &$5.94 \pm 0.14$ &$5.91 \pm 0.14$ &$6.00 \pm 0.47$ \\  \hline
$\chi^2/dof$ &$1.0/23$ &$14.9/85$ &$14.9/85$ &$11.1/85$ &$27.6/85$ \\  \hline

 \hline
\end{tabular}

\caption{Results of the fits of the AMS and PAMELA proton data with the various LIS models discussed in the text.}
\label{tabfit}

\end{table*}
 
\end{center}

\end{document}